\documentclass{article}
\usepackage{amssymb}
\usepackage{amsmath}
\usepackage{graphicx}

\begin{document}

\begin{center}
\Large\textbf{Interstitial diffusion of arsenic in silicon}
\\[2ex]
\normalsize
\end{center}

\begin{center}
\textbf{O. I. Velichko}

\bigskip


{\it E-mail address (Oleg Velichko):} velichkomail@gmail.com
\end{center}

\textit{Abstract.} The mechanism underlying the long-range
interstitial migration of nonequilibrium impurity interstitial
species was used to simulate arsenic redistribution in ion
implantation. An excellent agreement of the calculated arsenic
concentration profiles with experimental data allows one to assume
that the migration of nonequilibrium arsenic interstitial atoms
makes a significant contribution to the formation of a low
concentration region on thermal arsenic diffusion. The arsenic
concentration profile calculated for a temperature of 1050 Celsius
degrees within the framework of this assumption agrees well with
the experimental one. A number of parameters describing arsenic
diffusion at 1050 and 1108 Celsius degrees have been obtained.

\section{Introduction }
Recently, modeling arsenic thermal diffusion has been carried out
in \cite{Velichko-15} on the basis of the diffusion equation

\begin{equation} \label{DifEqEff}
\frac{\partial \, C}{\partial \, t} =\frac{\partial }{\partial \, x}
\left[D\left(\chi \right)\, h\, \left(C,C^{B} \right)\; \frac{\partial C}{\partial \, x} \, \right] .
\end{equation}

The different cases of concentration dependence $D^{C} (\chi )$ for the effective arsenic diffusivity $D\left(\chi \right)$ represented in the form

\begin{equation} \label{Deff}
D\left(\chi \right)\, =D_{i} D^{C} (\chi )=D_{i} \frac{1+\beta _{1} \chi +\beta _{2} \chi ^{2} }{1+\beta _{1} +\beta _{2} }
\end{equation}

\noindent have been investigated. It is supposed that

\begin{equation} \label{Chi}
\chi =\frac{\left(C-C^{B} \right)+\sqrt{\left(C-C^{B} \right)^{2}
+4n_{i}^{2} } }{2n_{i} } \, ,
\end{equation}

\begin{equation} \label{hMulti}
h\, \left(C,C^{B} \right)=1+\frac{C}{\sqrt{\left(C-C^{B}
\right)^{2} +4n_{i}^{2} } } \, ,
\end{equation}

\begin{equation} \label{Di}
D_{i} =D_{i}^{E\times } +D_{i}^{F\times } +D_{i}^{E-} +D_{i}^{E2-}
\, ,
\end{equation}

\begin{equation} \label{BT1EF}
\beta _{1} ={D_{i}^{E-}  \mathord{\left/{\vphantom{D_{i}^{E-}
\left(D_{i}^{E\times } +D_{i}^{F\times } \right)}}\right.\kern-
\nulldelimiterspace} \left(D_{i}^{E\times } +D_{i}^{F\times }
\right)} \, ,
\end{equation}

\begin{equation} \label{BT2EF}
\beta _{2} ={D_{i}^{E2-} \mathord{\left/{\vphantom{D_{i}^{E2-}
\left(D_{i}^{E\times } +D_{i}^{F\times }
\right)}}\right.\kern-\nulldelimiterspace} \left(D_{i}^{E\times }
+D_{i}^{F\times } \right)} \, .
\end{equation}

Here $C$ and $C^{B}$ are the concentrations of substitutionally
dissolved arsenic atoms and background impurity of the opposite
type of conductivity, respectively; $\chi $ is the concentration
of electrons , normalized to the intrinsic carrier concentration
$n_{i}$; $h\, \left(C,C^{B} \right)$ is the factor describing the
influence of the built-in electric field on the drift of charged
pairs; $D_{i}^{E\times} $, $D_{i}^{E1}$, and $D_{i}^{E2}$ are the
partial intrinsic diffusivities of dopant atoms due to the
interaction with neutral, singly, and doubly charged vacancies,
respectively; $D_{i}^{F\times}$ is the partial intrinsic
diffusivity of arsenic atoms due to the interaction with neutral
self-interstitials; $\beta _{1}$ and $\beta _{2}$ are the
empirical constants that describe the relative contribution of
singly and doubly charged intrinsic point defects to the impurity
diffusion.

The empirical constants $\beta_{1}$ and $\beta_{2}$ in expression
\eqref{Deff} can be found from the best fit to experimental data.
For example, in Ref. \cite{Martinez-Limia-08} a fitting routine
based on different experimental data was used and the temperature
dependences of the arsenic partial diffusivities $D_{i}^{\times }
$, $D_{i}^{-} $, $D_{i}^{2-}$due to neutral, singly, and doubly
charged intrinsic point defects were evaluated. The dependences
obtained in \cite{Martinez-Limia-08} allow one to calculate the
values of $\beta _{1}$ and $\beta _{2}$ for different temperatures
of thermal treatment. It was shown in \cite{Velichko-15} that the
concentration dependence of arsenic diffusivity obtained in
\cite{Martinez-Limia-08} provides excellent agreement with
experimental data for impurity concentration close to the value of
$n_{i} $. On the other hand, if the arsenic concentration
$C>>n_{i} $ and the concentration profile of electrically active
arsenic gets a ``box-like'' form (see Fig. 3 in
\cite{Velichko-15}), a difference from the experimental data is
observed in the local region where a strong decrease in the
impurity concentration begins. Therefore, for solving the problems
arising in modeling silicon doping with arsenic it seems more
convenient to use the concentration dependence proposed in
\cite{Tsai-80}, which is characterized by neglecting the
interaction of impurity atoms with doubly charged point defects.
Indeed, in this case, good agreement with experiment is observed
in the whole region of high impurity concentration and there is
only a small difference in the region of low arsenic
concentration  \cite{Velichko-15}. It was supposed in
\cite{Velichko-15} that this difference appears due to the direct
migration of arsenic interstitial atoms.

The goal of this work is to investigate the possibility of arsenic
interstitial diffusion and to obtain complete agreement of
modeling results with experimental data.

\section{Interstitial diffusion of arsenic atoms in
ion-implanted layers}

The assumption about the interstitial diffusion of arsenic atoms
during ion implantation was first made in \cite{Schwettmann-73}
with the purpose to explain the experimental data obtained. In
\cite{Schwettmann-73} the {\bf {\itshape p}}-type
Czochralski-grown silicon substrates of (111) orientation with a
resistivity of 5 -- 10 $\Omega$cm were implanted with As${}^{75}$
ions at 120 keV for doses of 1$\times$10${}^{12}$ --
1$\times$10${}^{16}$ ions/cm${}^{2}$. The average dose rate was
approximately 1 $\mu $A/cm$^{2}$. The implantation was carried out
at a room temperature with the beam at a 7$^{\circ}$ angle to the
silicon surface in an attempt to minimize channeling. After
implantation, the wafers were cleaned and annealed at 600 -- 900
$^{\circ}$C in nitrogen environment. The conductivity profile of
the implanted layer was obtained by using the incremental sheet
resistance technique. Irvin's well-known data were used to convert
the conductivity profile into the electrically active arsenic
concentration distribution. It was shown by a special experimental
procedure that the ``tail'' is formed during implantation and not
as a result of enhanced diffusion from the high concentration
region during the early stages of annealing. After annealing at
600 $^{\circ}$C, there was no evidence of an electrically active
arsenic ``tail''. When the annealing temperature was increased up
to 725 $^{\circ}$C, the tail became electrically active. The lack
of electrical activity at 600 $^{\circ}$C indicates that the
arsenic is not in its usual substitutional environment in the
``tail'' region. It means that vacancies are not the traps for
migrating interstitial species.

\begin{figure}[!ht]
\centering {
\begin{minipage}[!ht]{9.4 cm}
{\includegraphics[scale=0.8]{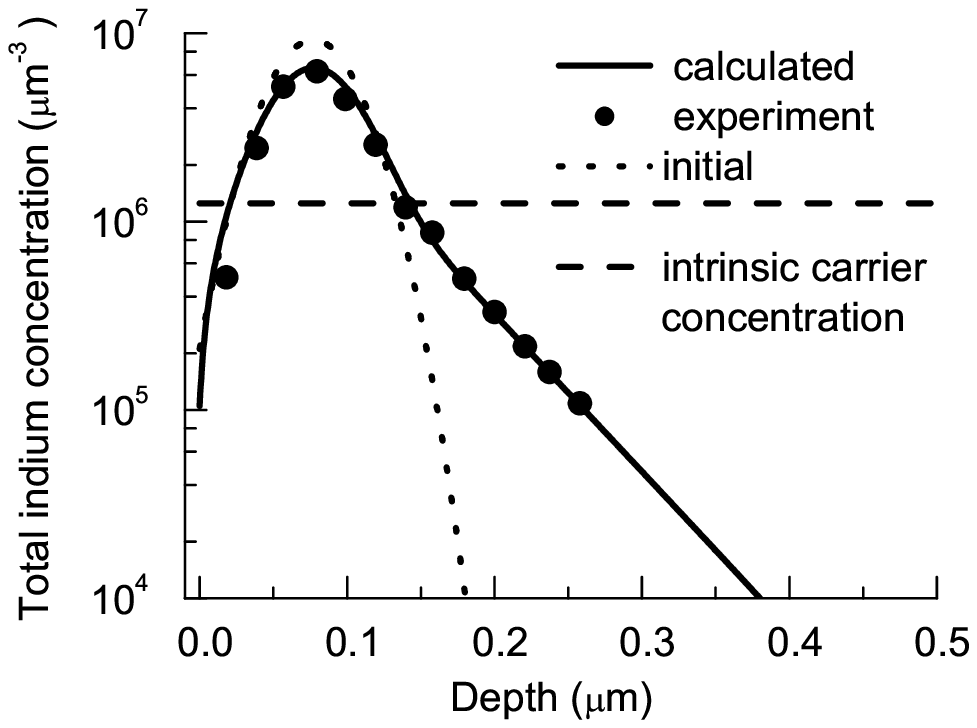}}
\end{minipage}}
\caption{Calculated concentration profile of electrically active
arsenic after annealing at a temperature of 725 $^{\circ}$C for 60
min. Implantation dose equals 5$\times$10${}^{1}$${}^{3}$
ions/cm${}^{2}$. Experimental data are taken from
\cite{Schwettmann-73}. The dotted curve is the Gaussian
distribution after implantation.} \label{fig:Sw-13}
\end{figure}

\begin{figure}[!ht]
\centering {
\begin{minipage}[!ht]{9.4 cm}
{\includegraphics[scale=0.8]{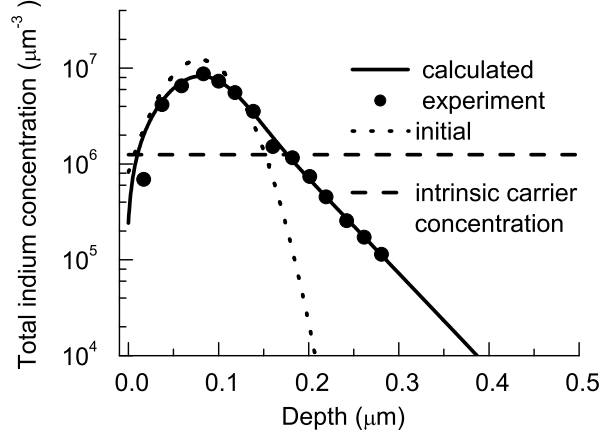}}
\end{minipage}}
\caption{Calculated concentration profile of electrically active
arsenic after annealing at a temperature of 725 $^{\circ}$C for 60
min. Implantation dose equals 1$\times$10${}^{14}$
ions/cm${}^{2}$. Experimental data are taken from
\cite{Schwettmann-73}. The dotted curve is the Gaussian
distribution after implantation.} \label{fig:Sw-14}
\end{figure}

In Figs.~\ref{fig:Sw-13} and ~\ref{fig:Sw-14} the arsenic
concentration profiles measured in \cite{Schwettmann-73} after
annealing at a temperature of 725 $^{\circ}$C for 60 minutes are
presented. The implantation doses are 5$\times$10${}^{13}$ and
1$\times$10${}^{14}$ ions/cm${}^{2}$, respectively. The chosen
values of the doses ensure the maximal impurity concentration
below a solubility limit of arsenic in silicon that is equal to
9.85$\times$10${}^{8}$ $\mu $m${}^{-3}$ for a temperature of 725
$^{\circ}$C and below an equilibrium electron concentration $n_{e}
$ = 9.32$\times$10${}^{7}$ $\mu $m${}^{-3}$ for this temperature
\cite{Solmi-01}. In Figs.~\ref{fig:Sw-13} and ~\ref{fig:Sw-14} we
also present the results of modeling the interstitial migration of
arsenic atoms during ion implantation. A model of interstitial
migration and an analytical solution of diffusion equation
proposed in \cite{Velichko-11} were used for simulation of
``tail'' formation. Taking into account the results of
 \cite{Schwettmann-73}, it was also supposed that the interstitial arsenic
species is trapped by uniformly distributed sinks but these sinks
are not vacancies. To provide a best fit to experimental data, the
Robin's boundary condition is imposed on the surface of a
semiconductor. Use of this boundary condition allows one to
describe the evaporation of arsenic interstitial species through
the surface.

The following values of the model parameters were used to provide
the best fit of the calculated arsenic concentration profile to
the experimental one.

\textbf{For implantation with the dose $Q$ =
5$\times$10${}^{1}$$^{3}$ ions/cm$^{-2}$ (Fig.~\ref{fig:Sw-13})}.

The parameters prescribing the initial distribution of implanted
arsenic are: $R_{p}$ = 0.077 $\mu$m, $\Delta R_{p}$ = 0.028
$\mu$m. The parameters specifying the process of interstitial
diffusion are: the average migration length of arsenic
interstitial species $l_{AI}$ = 0.052 $\mu$m, the fraction of the
arsenic atoms participating in the interstitial migration $p^{AI}$
= 50.1 \%.

\textbf{For implantation with the dose $Q$ =
1$\times$10${}^{1}$$^{4}$ ions/cm$^{-2}$ (Fig.~\ref{fig:Sw-14})}.

The parameters prescribing the initial distribution of implanted
arsenic are: $R_{p}$ = 0.079 $\mu$m, $\Delta R_{p}$ = 0.034 $\mu$m
The parameters specifying the process of interstitial diffusion
are: the average migration length of arsenic interstitial species
$l_{AI}$ = 0.044 $\mu$m, the fraction of the arsenic atoms
participating in the interstitial migration $p^{AI}$ = 70.4 \%.

It can be seen from Figs.~\ref{fig:Sw-13} and ~\ref{fig:Sw-14}
that there is excellent agreement between the calculated arsenic
concentration profiles and the electrically active arsenic
profiles measured in \cite{Schwettmann-73}. The agreement of
simulation results with experimental data indicates in favour of
the long-range interstitial migration of nonequilibrium arsenic
species. It is worth noting that the parameters prescribing the
initial distribution of implanted boron are approximately equal to
the values tabulated in \cite{Burenkov-86}: $R_{p}$ = 0.0712
$\mu$m, $\Delta R_{p}$ = 0.0248 $\mu$m, $Sk$ = 0.31, $R_{m}$ =
0.0677 $\mu$m. Here $Sk$ and $R_{m}$ are respectively the skewness
and the position of a maximum of impurity distribution as
implanted which is described by the Pearson type IV distribution
\cite{Burenkov-86}.

\section {Thermal arsenic diffusion}
\begin{figure}[!ht]
\centering {
\begin{minipage}[!ht]{9.4 cm}
{\includegraphics[scale=0.8]{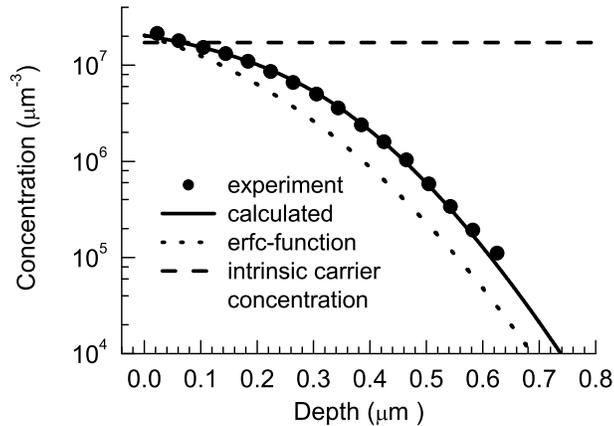}}
\end{minipage}}
\caption{Arsenic concentration profiles formed by thermal
diffusion from a constant source on the silicon surface. The solid
curve represents the arsenic concentration profile obtained by
numerical solution of the diffusion equation that takes account of
the concentration dependence of arsenic diffusivity and drift of
the charged species in the built-in electric field. Diffusion
temperature is 1108 $^{\circ}$C for 120 min. Filled circles are
the experimental data of \cite{Chiu-71}.} \label{fig:Chiu-Low}
\end{figure}

It was shown in \cite{Velichko-15} that the values of intrinsic
diffusivity $D_{i}$ = 2.55$\times$10${}^{-6}$ $\mu $m${}^{2}$/s
and the parameters $\beta _{1}$ = 611 and $\beta _{2}$ = 44.94
calculated from the expression given in \cite{Martinez-Limia-08}
for a temperature of 1108 $^{\circ}$C provides an agreement with
the experimental data \cite{Chiu-71} for impurity concentration
close to the value of $n_{i}$. In Ref. \cite{Chiu-71}, thermal
arsenic diffusion was carried out from a constant source on the
surface of a semiconductor. The arsenic concentration profile for
120 min thermal treatment measured in \cite{Chiu-71} by neutron
activation analysis is presented in Fig.~\ref{fig:Chiu-Low}. On
the other hand, the results of our simulation of arsenic diffusion
based on the model of \cite{Tsai-80}, which neglects doubly
charged point defects, give a similar agreement with the
experimental data \cite{Chiu-71}. The following parameters that
describe arsenic diffusion were used in this simulation: $D_{i} $
= 2.507$\times$10${}^{-}$${}^{6}$ $\mu $m${}^{2}$/s, $\beta _{1} $
= 100, and $\beta _{2} $ =0 \cite{Tsai-80}. This agreement is not
surprising because the maximal value of $\chi $ at the diffusion
temperature is equal to 1.78 and $\beta _{1} \chi $$>$$>$$\beta
_{2} \chi ^{2} $. Only for $\chi$ = 13.6 we have $\beta _{1} \chi
\approx \beta _{2} \chi ^{2}$. It means that for the diffusion
process under consideration a contribution of doubly charged point
defects is negligible. Then, due to the great values of $\beta
_{1}$ in the models \cite{Martinez-Limia-08} and \cite{Tsai-80},
the arsenic diffusivity can be approximated by the identical
expression

\noindent
\begin{equation} \label{DEffSimpl}
D\left(\chi \right)\, =D_{i} D^{C} (\chi )\approx D_{i} \, \chi  .
\end{equation}

\begin{figure}[!ht]
\centering {
\begin{minipage}[!ht]{9.4 cm}
{\includegraphics[scale=0.8]{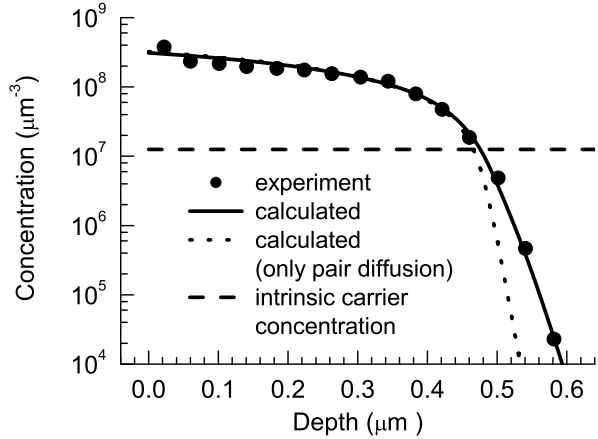}}
\end{minipage}}
\caption{Arsenic concentration profiles formed by thermal
diffusion from a constant source on the silicon surface. The
dotted curve represents the arsenic concentration profiles
obtained by numerical solution of the diffusion equation based on
the pair diffusion mechanism. The solid curve takes addition
account of the arsenic interstitial migration. Diffusion
temperature is 1050 $^{\circ}$C for 60 min. Filled circles are the
experimental data of \cite{Chiu-71}.} \label{fig:Pair-Int}
\end{figure}

In Fig.~\ref{fig:Chiu-Low} a similar numerical solution of the
diffusion equation \eqref{DifEqEff} for the doping process
investigated in \cite{Chiu-71} is also presented. Using a
variation of the parameters $D_{i} $ and $\beta _{1} $ ($\beta
_{2} $ = 0), we have achieved an ideal fit to the experimental
data of \cite{Chiu-71}. It is worth noting that the obtained
values of $D_{i} $ = 2.7$\times$10${}^{-6}$ $\mu $m${}^{2}$/s and
$\beta _{1} $ = 4, especially$\beta _{1} $, differ from the values
reported in \cite{Martinez-Limia-08} and \cite{Tsai-80}.
Therefore, it is interesting to investigate the arsenic diffusion
with a higher doping level. The results of simulation of high
concentration arsenic diffusion are presented in
Fig.~\ref{fig:Pair-Int}. The arsenic concentration profile
obtained in \cite{Chiu-71} for a diffusion temperature of 1050
$^{\circ}$C and duration of 60 minutes is used for comparison.

It can be seen from Fig.~\ref{fig:Pair-Int} that the arsenic
concentration near the surface is approximately equal to
3$\times$10${}^{8}$ $\mu $m${}^{-3}$ that results in $\chi \approx
{\rm 18.4}$. Thus, arsenic diffusion is characterized by a strong
nonlinear concentration dependence of effective diffusivity. On
the other hand, arsenic concentration is lower than the maximal
equilibrium electron concentration $n_{e}$ = 3.566$\times$10$^{8}$
$\mu $m$^{-3}$ for the annealing temperature under consideration
\cite{Solmi-01}. Therefore, it is possible to assume that arsenic
clustering does not play an important role in the retardation of
diffusion. The values of simulation parameters obtained from the
best fitting to the experimental concentration profile are: $D_{i}
$ = 5.8$\times$10${}^{-}$${}^{7}$ $\mu $m${}^{2}$/s and
$\beta_{1}$ = 4 ($\beta_{2}$ = 0). The impurity concentration at
the surface has been chosen equal to 3.2$\times$10$^{8}$ $\mu
$m$^{-3}$. It is worth noting that the obtained value of $D_{i}$
is practically the same as $D_{i}$ = 5.8134$\times$10${}^{-7}$
$\mu $m$^{2}$/s calculated for a temperature of 1050 $^{\circ}$C
from the expression proposed in \cite{Martinez-Limia-08}. On the
other hand, $\beta_{1}$ = 4 and $\beta_{2}$ = 0 used in simulation
differ significantly from the values reported in
\cite{Martinez-Limia-08}: $\beta _{1}$ = 238.9; and $\beta _{2}$ =
12.7. It can be seen from Fig. 4 that the calculated arsenic
concentration profile described by the dotted curve agrees well
with experimental data, except for the region of a low impurity
concentration. It is similar to the arsenic profile calculated in
\cite{Velichko-15} within a framework of the model of
\cite{Tsai-80} with $D_{i} $ = 5.537$\times$10${}^{-7}$ $\mu
$m${}^{2}$/s, $\beta _{1} $ = 100, and $\beta _{2} $ = 0.

Taking into account the results obtained, it is reasonable to
neglect the contribution of doubly charged point defects in
simulation of high concentration arsenic diffusion in silicon,
because in this case the calculated profile describes more
precisely the high concentration region and the region where a
strong decrease in the impurity concentration begins
\cite{Velichko-15}. To achieve the agreement with the experiment
in the low concentration region of arsenic distribution, we
suppose that generation of nonequilibrium interstitial arsenic
atoms can occur during high concentration arsenic diffusion.
Participating in the long-range interstitial migration, these
atoms form a low concentration region in the arsenic profile. For
example, such generation can occur in the region of the abrupt
fall in the arsenic concentration due to the stresses arising in
the layer between the highly doped and intrinsic silicon. It is
worth noting that in this layer intense dissociation of the
``impurity atoms
--- intrinsic point defect'' pairs also takes place, as follows
from the mass action law. After dissociation, the arsenic atoms
that previously formed pairs become substitutionally dissolved
again. However, a small fraction of arsenic atoms can occupy
interstitial position and participate in the long-range migration.
To take into account this additional impurity flux, the equation
for interstitial diffusion \cite{Velichko-11} can be used. It is
worth noting that thermal diffusion of arsenic occurs at
temperatures significantly greater than the temperature of ion
implantation in the experiments of \cite{Schwettmann-73}.
Therefore, in contrast to \cite{Schwettmann-73}, it is supposed in
this paper that arsenic interstitials become substitutionally
dissolved again via recombination with vacancies (the
Frank-Turnbull diffusion mechanism \cite{Frank-56}). Then, the
system of equations describing arsenic diffusion both by means of
the formation, migration, and dissociation of equilibrium
``impurity atoms --- intrinsic point defect'' pairs and due to the
long-range migration of nonequilibrium arsenic interstitials has
the form

\begin{equation} \label{DifEqEffInt}
\frac{\partial \, C}{\partial \, t} =\frac{\partial }{\partial \, x} \left[D\left(\chi \right)\, h\, \left(C,C^{B} \right)\; \frac{\partial C}{\partial \, x} \, \right]+\frac{C^{AI} (x,t)}{\tau ^{AI} } -G^{AI} (x,t) ,
\end{equation}

\begin{equation} \label{dAI}
d^{AI} \frac{\partial ^{2} C^{AI} }{\partial \, x^{2} } -\frac{C^{AI} (x,t)}{\tau ^{AI} } +G^{AI} (x,t)=0 .
\end{equation}

Here $C^{AI} $ is the concentration of nonequilibrium interstitial
arsenic atoms; $d^{AI}$ and $\tau ^{AI}$ are the diffusivity and
average lifetime of these nonequilibrium interstitial atoms,
respectively; $G^{AI}$ is the generation rate of arsenic
interstitials per unit volume of the semiconductor. We use the
stationary diffusion equation for interstitial arsenic atoms in
view of their large average migration length $l_{AI} =\sqrt{d^{AI}
\tau ^{AI}}$ and small lifetime ($\tau ^{AI} <<\tau _{p} $), where
$\tau _{p}$ is the duration of thermal treatment.

A numerical solution of the system of diffusion equations
\eqref{DifEqEffInt} and \eqref{dAI} for the process investigated
in \cite{Chiu-71} is presented in Fig.~\ref{fig:Pair-Int} by a
solid curve. It can be seen from Fig.~\ref{fig:Pair-Int} that
account of the migration of nonequilibrium arsenic interstitials
allows one to explain the formation of the ``tail'' in the low
concentration region and to achieve complete agreement of the
arsenic concentration profile with the experimental one. The
values of the simulation parameters that provide the best fit to
the experimental concentration profile are: $D_{i}$ =
6.2$\times$10$^{-}$$^{7}$ $\mu $m$^{2}$/s and $\beta_{1}$ = 4
($\beta_{2}$ = 0). The impurity concentration at the surface
$C_{S} $ has been chosen equal to 3.1$\times$10$^{8}$
$\mu$m$^{-3}$ and the average migration length of nonequilibrium
arsenic interstitials $l_{AI}$ = 0.013 $\mu$m. It can be seen from
Fig.~\ref{fig:Pair-Int} that the calculated arsenic concentration
profile described by the solid curve agrees well with experimental
data over the whole diffusion zone. It is worth noting that full
agreement with the measured concentration profile has been
achieved using the assumption of the long-range migration of
nonequilibrium arsenic interstitials. This assumption is based on
the results of the previous simulation of arsenic diffusion during
ion implantation. On the other hand, there can be another way to
describe the low concentration ``tail'' region of arsenic
distribution. For example, it was shown in \cite{Velichko-86} that
in the low concentration region a condition of a local
thermodynamic equilibrium between the substitutionally dissolved
impurity atoms, intrinsic point defects, and ``impurity atom ---
intrinsic point defect'' pairs can be broken up. In this case, the
nonequilibrium pairs will be playing the role of nonequilibrium
arsenic interstitials. Moreover, it was shown in
\cite{Velichko-87} that a similar ``tail'' region can be formed in
the case of diffusion of equilibrium ``impurity atom --- intrinsic
point defect'' pairs if the nonuniform distribution of intrinsic
point defects in the neutral charge state arises. Such nonuniform
distribution can be formed due to the intense dissociation of
``impurity atoms --- intrinsic point defect'' pairs in the region
of abrupt fall of arsenic concentration.

\section{Conclusions}

On the basis of the mechanism of the long-range interstitial
migration of nonequilibrium impurity interstitial species,
simulation of arsenic concentration profiles measured in
\cite{Schwettmann-73} after implantation and annealing of silicon
substrates at a temperature of 725 $^{\circ}$C for 60 minutes has
been carried out. The arsenic concentration profiles calculated
for ion implantation with an energy of 120 keV and doses of
5$\times$10${}^{13}$ and 1$\times$10${}^{1}$${}^{4}$ cm${}^{-2}$
agree well with experimental ones. Thus, the simulation results
obtained confirm the assumption of \cite{Schwettmann-73} that the
``tail'' in the low concentration regions of arsenic profiles is
formed due to the interstitial migration of arsenic species during
ion implantation. The average migration length of arsenic
interstitial species obtained from the best fit to the
experimental concentration profiles are equal to 0.052 and 0.044
$\mu$m for implantation with doses of 5$\times$10${}^{13}$ and
1$\times$10${}^{1}$${}^{4}$ cm${}^{-2}$, respectively.

Based on the results obtained, simulation of thermal arsenic
diffusion has been carried out. It is shown that assumption of the
long-range migration of nonequilibrium arsenic interstitial atoms
allows one to explain the formation of the ``tail'' in the low
concentration region for the case of high concentration arsenic
diffusion. Complete agreement of arsenic concentration profile
with the experimental one \cite{Chiu-71} has been achieved over
the whole diffusion zone. The average migration length of
nonequilibrium arsenic interstitials obtained from the best fit to
the experimental concentration profile is equal to 0.013 $\mu $m.
It is worth noting that the empirical parameter $\beta _{1} $ that
describes the relative contribution of singly charged intrinsic
point defects to the impurity diffusion is equal to the same value
for diffusion at temperatures of 1108 and 1050 $^{\circ}$C. The
value of $\beta _{1}$ = 4 obtained from the best fit to the
experimental arsenic profiles differs significantly from the
values used in \cite{Martinez-Limia-08} and \cite{Tsai-80}.

\end{document}